\documentclass[structabstract]{aa}  
\usepackage{graphicx}
\usepackage{txfonts}
\usepackage{natbib}
\usepackage[breaklinks=false]{hyperref}
\newcommand\cst{{\ensuremath{\mathrm{cst}}}}
\newcommand\dif{{\ensuremath{\mathrm{d}}}}
\newcommand\dln{{\ensuremath{\mathrm{dln}}}}
\newcommand\ds{{\ensuremath{\dif s}}}
\newcommand\dt{{\ensuremath{\dif t}}}
\newcommand\dr{{\ensuremath{\dif r}}}
\newcommand\dP{{\ensuremath{\dif P}}}
\newcommand\dMr{{\ensuremath{\dif M_r}}}

\newcommand\dphi{{\ensuremath{\dif\varphi}}}
\newcommand\dpsi{{\ensuremath{\dif\psi}}}
\newcommand\dxi{{\ensuremath{\dif\xi}}}
\newcommand\Ms{{\ensuremath{\mathrm{M}_{\odot}}}}

\newcommand\kb{\ensuremath{k_{\rm B}}}
\newcommand\mh{\ensuremath{m_{\rm H}}}
\newcommand\kmu{{\kb\over\mu\mh}}
\newcommand\asb{\ensuremath{a_{\rm SB}}}
\newcommand\rhc{\ensuremath{\rho_c}}
\newcommand{\Mpy}{\Ms\,{\rm yr}{\ensuremath{^{-1}}}}
\newcommand{\dm}{\ensuremath{\dot M}}
\newcommand{\gva}{{\sc genec}}
\newcommand{\gaz}{{\ensuremath{_{\rm gas}}}}
\newcommand{\rad}{{\ensuremath{_{\rm rad}}}}
\newcommand{\pgaz}{{\ensuremath{P\gaz}}}

\newcommand{\srad}{{\ensuremath{s\rad}}}
\newcommand{\xisurf}{{\ensuremath{\xi_{\rm surf}}}}
\newcommand{\xicore}{{\ensuremath{\xi_{\rm core}}}}
\newcommand{\phicore}{{\ensuremath{\varphi_{\rm core}}}}
\newcommand{\phisurf}{{\ensuremath{\varphi_{\rm surf}}}}
\newcommand{\phimin}{{\ensuremath{\varphi_{\rm min}}}}
\newcommand{\phimax}{{\ensuremath{\varphi_{\rm max}}}}
\newcommand{\Mcore}{{\ensuremath{M_{\rm core}}}}
\newcommand{\crit}{{\ensuremath{\left.{\beta\over\sigma}\right\vert_{\rm crit}}}}

\usepackage{color}

\begin{document}

\title{General-relativistic instability in hylotropic supermassive stars}
%\titlerunning{.}

\author{L. Haemmerl\'e}
\authorrunning{Haemmerl\'e}

\institute{D\'epartement d'Astronomie, Universit\'e de Gen\`eve, chemin des Maillettes 51, CH-1290 Versoix, Switzerland}

%\date{Received ; accepted }

% \abstract{}{}{}{}{} 
% 5 {} token are mandatory
 
\abstract
{The formation of supermassive black holes by direct collapse would imply the existence of supermassive stars (SMSs)
and their collapse through the general-relativistic (GR) instability into massive black hole seeds.
However, the final mass of SMSs is weakly constrained by existing models,
in spite of the importance this quantity plays in the consistency of the direct collapse scenario.}
{We estimate the final masses of spherical SMSs in the whole parameter space relevant for these objects.}
{We build analytical stellar structures (hylotropes) that mimic existing numerical SMS models accounting for full stellar evolution with rapid accretion.
From these hydrostatic structures, we determine ab initio the conditions for GR instability,
and compare the results with the predictions of full stellar evolution.}
{We show that hylotropic models predict the onset of GR instability with high precision.
The mass of the convective core appears as a decisive quantity.
The lower it is, the larger is the total mass required for GR instability.
Typical conditions for GR instability are a total mass $\gtrsim10^5$~\Ms\ with a core mass $\gtrsim10^4$~\Ms.
If the core mass remains below $10^4$~\Ms, total masses in excess of $10^6-10^7$~\Ms\ can be reached.}
{Our results confirm that spherical SMSs forming in primordial, atomically cooled haloes collapse at masses below 500~000~\Ms.
On the other hand, accretion rates in excess of 1000 \Mpy, leading to final stellar masses $\gtrsim10^6$ \Ms,
are required for massive black hole formation in metal-rich gas.
Thus, the different channels of direct collapse imply distinct final masses for the progenitor of the black hole seed.}

   %\keywords{}
 
\maketitle
%
%________________________________________________________________

\section{Introduction}
\label{sec-in}

Black hole seeds with masses $\gtrsim10^4$ \Ms, born via direct collapse during the formation process of their host galaxies,
appear as a natural explanation for the existence of the most massive black holes observed at high redshift
(e.g.~\citealt{rees1978,rees1984,volonteri2008,woods2019,haemmerle2020a}).
The direct progenitors of these objects would be supermassive stars (SMSs) growing by accretion at rates $\dm>0.1$~\Mpy\
until they collapse through the general-relativistic (GR) instability \citep{chandrasekhar1964}.
SMSs could form at the centre of protogalaxies provided the absence of molecular hydrogen
(e.g.~\citealt{haiman1997a,omukai2001a,bromm2003b,dijkstra2008}).
Hydrodynamical simulations show that the core of such primordial, atomically cooled haloes would collapse with negligible fragmentation
at inflows of 0.1 -- 10~\Mpy\ below parsec scales \citep{latif2013e}.
Another efficient mechanism for the formation of massive black hole seeds is the merger of massive, gas-rich galaxies \citep{mayer2010,mayer2015,mayer2019}.
In addition to allow for inflows as high as $10^5$ \Mpy\ at parsec scales,
this formation channel does not require primordial chemical composition, and metallicities up to solar are expected.
Thus, the progenitors of massive black hole seeds could as well be Population I (Pop I) or Population III (Pop III) SMSs.

The evolution of SMSs under rapid accretion has been addressed in several works over the last decade
\citep{begelman2010,schleicher2013,hosokawa2013,sakurai2015,umeda2016,woods2017,haemmerle2018a,haemmerle2018b,haemmerle2019a,haemmerle2019c}.
For $\dm\gtrsim0.01$ \Mpy, they evolve as red supergiant protostars \citep{hosokawa2013},
with most of their mass contained in a radiative envelope \citep{begelman2010}.
As nearly-Eddingon stars, their pressure support and entropy content are dominated by radiation so that, above a given mass,
small post-Newtonian corrections, negligible regarding the hydrostatic structure,
prevent the pressure to restaure the fragile equilibrium against adiabatic perturbations \citep{chandrasekhar1964}.
For Pop III SMSs, the GR instability leads in general to the formation of a black hole seed of nearly the same mass as the progenitor
\citep{fricke1973,shapiro1979,uchida2017},
but once metals are included, thermonuclear explosions could prevent black hole formation
\citep{appenzeller1972a,appenzeller1972b,fuller1986,montero2012}.
Such outcomes depend sensitively on the mass of the SMS at collapse, which is weakly constrained by the models \citep{woods2019},
especially in the Pop I case.

In a previous work \citep{haemmerle2020b},
we showed how the onset point of the GR instability can be determined with high accuracy from hydrostatic stellar models.
However, the numerical estimates of the final masses suffer from the lack of such models up to the largest masses and accretion rates.
The exploration of the parameter space with full stellar evolution
is strongly limited by the numerical instability of stellar evolution codes when rapid accretion is included.
In the present work, we circumvent this difficulty with the help of simplified analytical hydrostatic structures,
built to match the structures obtained with full stellar evolution calculations accounting for rapid accretion.
Hylotropes have been proposed by \cite{begelman2010} to describe the envelope of the 'quasi-star' model, where a black hole grows at the centre of a SMS.
Numerical models accounting for full stellar evolution showed that this class of structures describes with a high precision
the actual envelope of SMSs in the extreme accretion regime \citep{haemmerle2019c}.
Here, we apply the method presented in \cite{haemmerle2020b} to the full class of hylotropes,
and discuss the implications regarding the final masses of SMSs in the various versions of direct collapse.

The main properties of rapidly accreting SMSs are reviewed in section \ref{sec-sms},
with an emphasise on the physical motivations for the use of hylotropes.
The methods followed to build the hylotropic models and to determine their stability are described in sections \ref{sec-hylo} and \ref{sec-gr}, respectively.
The results are presented in section \ref{sec-res} and discussed in section \ref{sec-dis}.
We summarise our conclusions in section \ref{sec-out}.

\section{Method}
\label{sec-meth}

\subsection{The structure of rapidly accreting SMSs}
\label{sec-sms}

Relativistic corrections remain always small in SMSs,
and regarding their equilibrium properties these stars are well described by the classical equations of stellar structure
(e.g.~\citealt{hoyle1963a,fuller1986,hosokawa2013,woods2017,haemmerle2018a}).
The equation of state is given by the sum of gas and radiation pressure
\begin{eqnarray}
P=\kmu\rho T+{1\over3}\asb T^4,
\label{eq-eos}\end{eqnarray}
where $P$ is the total pressure, $\rho$ the mass-density, $T$ the temperature, $\mu\mh$ the average mass of the free particles of gas,
\kb\ the Boltzmann constant and \asb\ the Stefan-Boltzmann constant.
SMSs are always close to the Eddington limit, with an entropy \srad\ given by radiation and a ratio of gas to total pressure of the order of a percent:
\begin{eqnarray}
\beta:={\pgaz\over P}={1\over1+{\mu\mh\over3\kb}{\asb T^3\over\rho}}={1\over1+{\mu\mh\over4\kb}\srad}	\ll1
\label{eq-beta}\end{eqnarray}
Pressure can be expressed as a function of $\rho$ and $\beta$:
\begin{eqnarray}
P&=&\left({3\over\asb}\right)^{1/3}\left(\kmu\right)^{4/3}\left({1-\beta\over\beta^4}\right)^{1/3}\rho^{4/3}	\nonumber\\
&\simeq&\left({3\over\asb}\right)^{1/3}\left(\kmu\right)^{4/3}\beta^{-4/3}\rho^{4/3}
\label{eq-eosbeta}\end{eqnarray}
Due to convection driven by H-burning, the core of SMSs is isentropic ($\beta=\cst=:\beta_c$),
and therefore, according to equation (\ref{eq-eosbeta}), a polytrope of index $n=3$:
\begin{eqnarray}
P=K\rho^{4/3},\qquad
K=\left({3\over\asb}\right)^{1/3}\left(\kmu\right)^{4/3}\left({1-\beta\over\beta^4}\right)^{1/3}=\cst
\label{eq-poly}\end{eqnarray}
On the other hand, SMSs accrete at rates $\dm>0.1$ \Mpy, for which the accretion time $M/\dm$ is shorter than the thermal timescale
\citep{hosokawa2013,haemmerle2018a,haemmerle2019c}.
As a consequence, the envelope is not relaxed thermally, and keeps a higher entropy than the core, inhibiting convection.
For rates $\dm>10$ \Mpy, thermal processes become completely negligible
and the radiative envelope evolves adiabatically for most of its mass \citep{haemmerle2019c}.
In this case, the entropy profile builds homologously, which implies the following dependence on the mass coordinate $M_r$ ($r$ is the radial coordinate):
\begin{equation}
s\propto{T^3\over\rho}\propto{P^{3/4}\over\rho}\propto\left({M_r^2\over r^4}\right)^{3/4}{r^3\over M_r}=M_r^{1/2}
\quad\Longrightarrow\ \beta\propto M_r^{-1/2}
\label{eq-shylo}\end{equation}
where we used equation (\ref{eq-beta}) to express $\beta$.
With this scaling law, equation (\ref{eq-eosbeta}) gives a hylotrope \citep{begelman2010}:
\begin{eqnarray}
P=AM_r^{2/3}\rho^{4/3}
\label{eq-hylo}\end{eqnarray}
where $A$ is a constant.
The hylotropic law (\ref{eq-hylo}) switches into a polytropic law (\ref{eq-poly}) at the limit of the convective core,
i.e. at a mass coordinate $M_r=\Mcore$.
Continuity of pressure at the interface imposes
\begin{equation}
K=AM_{\rm core}^{2/3}
\label{eq-match}\end{equation}
Thus, the structure of a maximally accreting SMS satisfies
\begin{eqnarray}
P&=&\left\{\begin{array}{ll}
K\rho^{4/3}								\qquad&{\rm if}\ M_r<\Mcore\\
K\left({M_r\over\Mcore}\right)^{2/3}\rho^{4/3}		\qquad&{\rm if}\ M_r>\Mcore
\end{array}\right.
\label{eq-smshylo}\\
\beta&=&\left\{\begin{array}{ll}
\beta_c								\quad\ \qquad&{\rm if}\ M_r<\Mcore	\\
\beta_c\left({M_r\over\Mcore}\right)^{-1/2}		\quad\ \qquad&{\rm if}\ M_r>\Mcore
\end{array}\right.
\label{eq-smsbeta}\end{eqnarray}
with
\begin{equation}
K=\left({3\over\asb}\right)^{1/3}\left({\kb\over\mu\mh}\right)^{4/3}\left({1-\beta_c\over\beta_c^4}\right)^{1/3}
\label{eq-K}\end{equation}

\begin{figure}\begin{center}
\includegraphics[width=.45\textwidth]{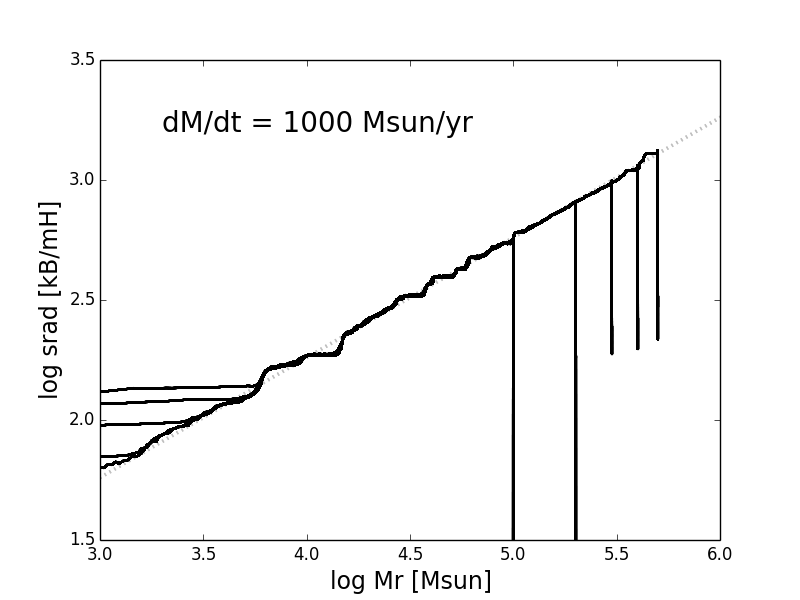}
\caption{Entropy profiles of the \gva\ model at zero metallicity and $\dm=1000$ \Mpy, at masses $M=1-2-3-4-5\times10^5$ \Ms.
The grey dotted line indicates the power-law $s=1.82{\kb\over\mh}(M_r/\Ms)^{1/2}$.}
\label{fig-genec1000}\end{center}\end{figure}

\begin{figure}\begin{center}
\includegraphics[width=.45\textwidth]{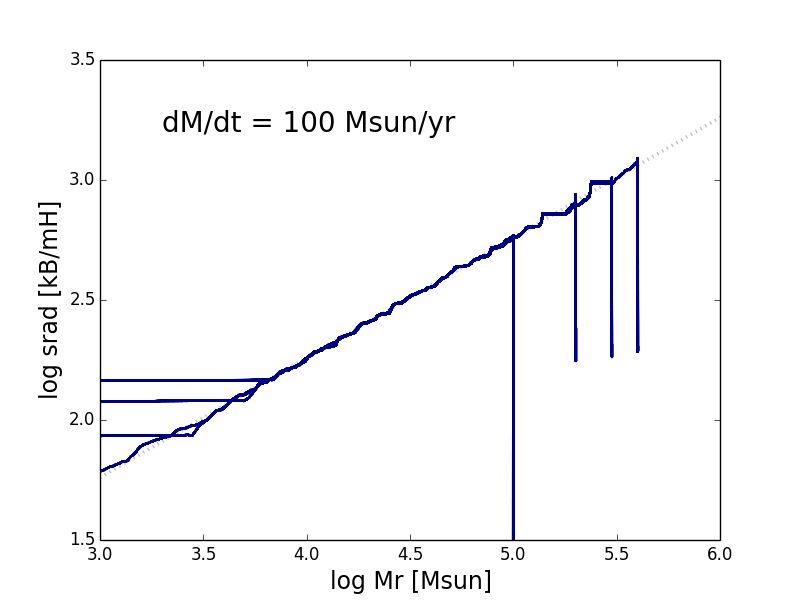}
\caption{Same as figure \ref{fig-genec1000} for $\dm=100$ \Mpy\ and $M=1-2-3-4\times10^5$ \Ms.}
\label{fig-genec100}\end{center}\end{figure}

\begin{figure}\begin{center}
\includegraphics[width=.45\textwidth]{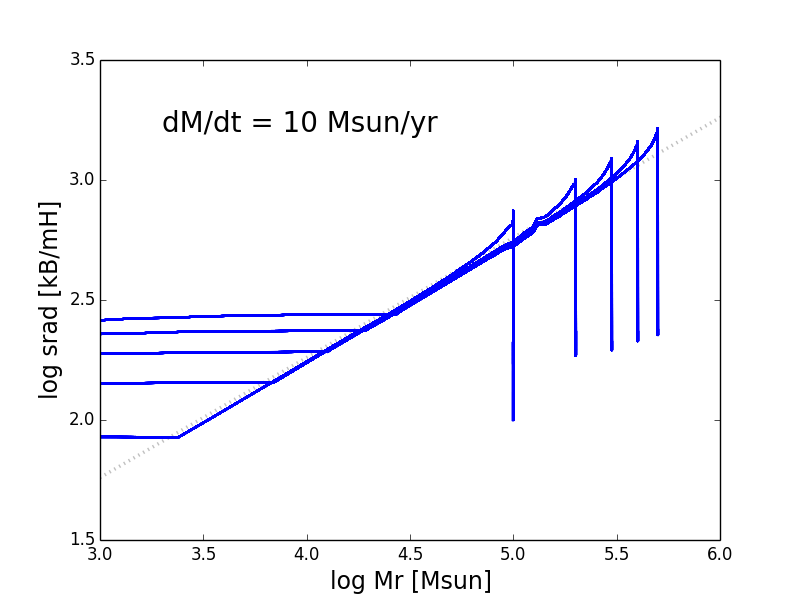}
\caption{Same as figure \ref{fig-genec1000} for $\dm=10$ \Mpy\ and $M=1-2-3-4-5\times10^5$ \Ms.}
\label{fig-genec10}\end{center}\end{figure}

\begin{figure}\begin{center}
\includegraphics[width=.45\textwidth]{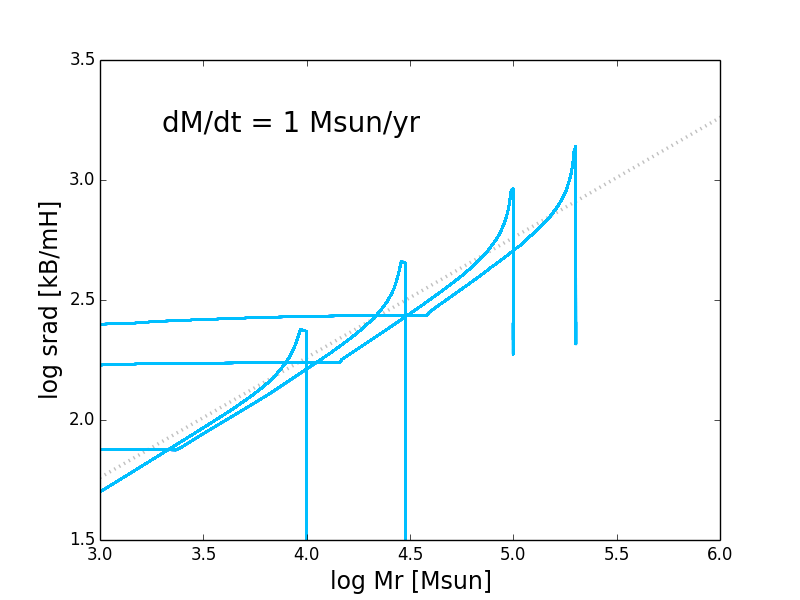}
\caption{Same as figure \ref{fig-genec1000} for $\dm=1$ \Mpy\ and $M=0.1-0.3-1-2\times10^5$ \Ms.}
\label{fig-genec1}\end{center}\end{figure}

The entropy profiles of the \gva\ models of \cite{haemmerle2018a,haemmerle2019c} at zero metallicity and $\dm=1-1000$ \Mpy\
are shown in figures \ref{fig-genec1000}-\ref{fig-genec100}-\ref{fig-genec10}-\ref{fig-genec1}.
These profiles are similar in all respects to those found with other codes \citep{hosokawa2013,umeda2016}.
The limit between the polytropic, convective core and the hylotropic, radiative envelope appears clearly.
An additional feature is visible, the nearly vertical outter end of the profiles.
The layers newly accreted gain entropy as they are incorporated in the stellar interior, which triggers convection.
But it impacts only the outter percent of the stellar mass, where the density is the lowest, and this feature can be neglected for our purpose.
We see that for rates $\dm\geq100$ \Mpy\ the successive entropy profiles match each other in the envelope, following the hylotropic law (\ref{eq-shylo}).
For 10 \Mpy, the external layers depart slightly from the hylotropic law.
For lower rates the hylotropic approximation is broken in the whole envelope.

\subsection{Hylotropic structures}
\label{sec-hylo}

Provided a constraint on pressure $P=f(\rho,M_r)$, classical spherical stellar structures are fully determined by hydrostatic equilibrium and continuity of mass:
\begin{eqnarray}
{\dP\over\dr}&=&-\rho{GM_r\over r^2}	\label{eq-hydro}\\
{\dMr\over\dr}&=&4\pi r^2\rho			\label{eq-continu}
\end{eqnarray}
where $G$ is the gravitational constant.
We introduce the dimensionless functions $\xi$, $\theta$, $\varphi$ and $\psi$:
\begin{eqnarray}
\xi=\alpha r,\quad\rho=\rho_c\theta^3,\quad\varphi={\alpha^3\over4\pi\rhc}M_r,\quad P=P_c\psi
\label{eq-adim}\end{eqnarray}
where $\rhc$ and $P_c$ are the central mass-density and pressure, and $\alpha$ is related to them by
\begin{eqnarray}
\alpha^2={\pi G\rhc^2\over P_c}
\label{eq-alpha}\end{eqnarray}
The boundary conditions are given by
\begin{eqnarray}
r=0	&:&\ 		\xi=0,\ \ \theta=\psi=1,\ \ \varphi=0						\label{eq-centre}\\
r=R	&:&\ 		\xi=\alpha R=:\xisurf,\ \ \theta=\psi=0,\ \ \varphi={\alpha^3M\over4\pi\rhc}=:\phisurf	\label{eq-surf}
\end{eqnarray}
where $M$ and $R$ are the total mass and radius of the star,
with \phisurf\ and \xisurf\ their dimensionless equivalents.
With these dimensionless functions, equations (\ref{eq-hydro}-\ref{eq-continu}) become
\begin{eqnarray}
{\dpsi\over\dxi}&=&-4{\varphi\theta^3\over\xi^2}	\label{eq-hydroadim}\\
{\dphi\over\dxi}&=&\xi^2\theta^3			\label{eq-continuadim}
\end{eqnarray}
The constraint on pressure reduces to a function $\psi(\theta,\varphi)$ that closes equations (\ref{eq-hydroadim}-\ref{eq-continuadim}).
Translated in the dimensionless functions (\ref{eq-adim}), equations (\ref{eq-smshylo}-\ref{eq-smsbeta}) give
\begin{equation}
\psi=\left\{\begin{array}{ll}
\theta^4								\qquad&{\rm if}\ \varphi<\phicore	\\
\left({\varphi\over\phicore}\right)^{2/3}\theta^4	\qquad&{\rm if}\ \varphi>\phicore
\end{array}\right.
\label{eq-hyloadim}\end{equation}
\begin{equation}
\beta=\left\{\begin{array}{ll}
\beta_c								\qquad&{\rm if}\ \varphi<\phicore\\
\beta_c\left({\varphi\over\phicore}\right)^{-1/2}		\qquad&{\rm if}\ \varphi>\phicore
\end{array}\right.
\label{eq-betaadim}\end{equation}
where
\begin{equation}
\phicore:={\alpha^3\over4\pi\rhc}\Mcore
\label{eq-phicore}\end{equation}
is the dimensionless mass of the convective core, that remains the only free parameter in the equation system.

Hylotropic structures can be built by numerical integration of (\ref{eq-hydroadim}-\ref{eq-continuadim}) with constraint (\ref{eq-hyloadim}),
starting from central conditions (\ref{eq-centre}).
For the numerical integration of the hylotropic envelope ($\varphi>\phicore$), we follow the more suitable Lagrangian formulation of \cite{begelman2010}
who defines $w=\dln\xi/\dln\varphi$.
In terms of this variable, equations (\ref{eq-hydro}-\ref{eq-continu}) with (\ref{eq-hylo}) lead to
\begin{eqnarray}
{\dln\varphi\over\dln w}&=&{1\over{3\over2}-3w+3\varphi_{\rm core}^{2/3}w^{4/3}}	\label{eq-w1}\\
{\dln\xi\over\dln w}&=&{w\over{3\over2}-3w+3\varphi_{\rm core}^{2/3}w^{4/3}}		\label{eq-w2}
\end{eqnarray}
The functions $\theta$, $\psi$ and $\beta$ are then obtained by equations (\ref{eq-continuadim}-\ref{eq-hyloadim}-\ref{eq-betaadim}).
Since $w\to\infty$ when $\rho\to0$, integration with respect to $w$ never reaches the surface.
We stop the integration when $\theta<10^{-8}$.
We build in this way a series of hylotropes with \phicore\ ranging in the interval
\begin{equation}
\phimin:=0.459\ <\ \phicore\ <\ 2.02=:\phimax
\label{eq-phiminmax}\end{equation}
by steps of 0.05 -- 0.3.
The lower limit of this interval is imposed by the condition of gravitational binding \citep{begelman2010},
while the upper limit corresponds to polytropes, i.e. $\phicore=\phisurf$.

The dimensional quantities can then be derived from the dimensionless ones.
The polytropic law (\ref{eq-poly}) in the core implies $P_c=K\rho_c^{4/3}$.
With the definition (\ref{eq-alpha}) of $\alpha$ and the expression (\ref{eq-K}) of $K$, it gives a mass-scale
\begin{equation}
{4\pi\rhc\over\alpha^3}={4\over\sqrt{\pi}}\left({K\over G}\right)^{3/2}={4\over G^{3/2}}\left({3\over\pi\asb}\right)^{1/2}\left({\kb\over\mu\mh}\right)^2\left({1-\beta_c\over\beta_c^4}\right)^{1/2}
\label{eq-mscale}\end{equation}
We see that the mass-scale is fully determined by $\beta_c$ and the chemical composition.
The choice of the central temperature, imposed by H-burning, determines $\rho_c$ through equation (\ref{eq-beta}),
which defines the length-scale $\alpha^{-1}$ through the mass-scale (\ref{eq-mscale}).

\subsection{GR instability in the post-Newtonian + Eddington limit}
\label{sec-gr}

We showed in \cite{haemmerle2020b} that the onset point of the GR instability can be determined with high accuracy from numerical hydrostatic structures
by a simple application of the relativisitic equation of adiabatic pulsation of \cite{chandrasekhar1964} in the following form:
\begin{eqnarray}
{\omega^2\over c^2}I_0=\sum_{i=1}^4I_i
\label{eq-chandra}\end{eqnarray}
with
\begin{eqnarray}
I_0&=&\int_0^Re^{a+3b}(P+\rho c^2)r^4\dr					\label{eq-I0}\\
I_1&=&9\int_0^Re^{3a+b}\left(\Gamma_1-{4\over3}\right)Pr^2\dr	\label{eq-I1}\\
I_2&=&-12\int_0^Re^{3a+b}\left(a'+{b'\over3}\right)Pr^3\dr			\label{eq-I2}\\
I_3&=&{8\pi G\over c^4}\int_0^Re^{3(a+b)}P(P+\rho c^2)r^4\dr		\label{eq-I3}\\
I_4&=&-\int_0^Re^{3a+b}{P'^2\over P+\rho c^2}r^4\dr			\label{eq-I4}
\end{eqnarray}
where $'$ indicates the derivatives with respect to the radial coordinate $r$ ($r=R$ at the surface),
$P$ is the pressure, $\rho$ the density of relativistic mass,
$c$ the speed of light, $G$ the gravitational constant, $\Gamma_1$ the first adiabatic exponent,
and $a$ and $b$ the coefficients of the metric:
\begin{equation}
\ds^2=-e^{2a}(c\dt)^2+e^{2b}\dr^2+r^2\dif\Omega^2
\label{eq-ds}\end{equation}
Einstein's equations lead to the following expressions for a spherical, static metric:
\begin{eqnarray}
a'&=&{GM_r\over r^2c^2}{1+{4\pi r^3\over M_rc^2}P\over1-{2GM_r\over rc^2}}	,\qquad
e^{2a(R)}=1-{2GM\over Rc^2}		\label{eq-a}\\
e^{-2b}&=&1-{2GM_r\over rc^2}	\label{eq-b}
\end{eqnarray}
The relativistic mass-coordinates $M_r$ ($M=M_R$) is related to the other quantities by equation (\ref{eq-continu})
with relativistic-mass instead of rest-mass in $M_r$ and $\rho$.

The condition for GR instability is
\begin{eqnarray}
\sum_{i=1}^4I_i<0
\label{eq-inst}\end{eqnarray}
Here, we formulate this condition for the special case of post-Newtonian stars near the Eddington limit.
To that aim, we reexpress integrals $I_i$ in the dimensionless functions (\ref{eq-adim}).
Following \cite{tooper1964a}, we define also
\begin{equation}
\sigma:={P_c\over\rhc c^2}={\pi G\rhc\over\alpha^2c^2}
\label{eq-sigma}\end{equation}
which represents the dimensionless parameter for the relativistic correction, since compactness is given by
\begin{equation}
{2GM_r\over rc^2}=8\sigma{\varphi\over\xi}
\label{eq-compact}\end{equation}
In these dimensionless quantities, integrals (\ref{eq-I1}-\ref{eq-I2}-\ref{eq-I3}-\ref{eq-I4}) read
\begin{eqnarray}
I_1&=&9{P_c\over\alpha^3}\int e^{3a+b}\left(\Gamma_1-{4\over3}\right)\psi \xi^2\dxi											\label{eq-I1adim}\\
I_2&=&-12{P_c\over\alpha^3}\int e^{3a+b}\left({\dif a\over\dxi}+{1\over3}{\dif b\over\dxi}\right)\psi\xi^3\dxi							\label{eq-I2adim}\\
I_3&=&8\sigma{P_c\over\alpha^3}\int e^{3(a+b)}\left(1+\sigma{\psi\over\theta^3}\right)\psi\theta^3\xi^4\dxi							\label{eq-I3adim}\\
I_4&=&-\sigma{P_c\over\alpha^3}\int e^{3a+b}{\left({\dpsi\over\dxi}\right)^2\over1+\sigma{\psi\over\theta^3}}{\xi^4\dxi\over\theta^3}	\label{eq-I4adim}
\end{eqnarray}
where the integration is made over the interval $0<\xi<\xisurf$.
Each integral has the same dimensional factor $P_c/\alpha^3$ in front, that can be extracted from the sum and cancelled in inequality (\ref{eq-inst}).

In the Eddington limit, the ratio of gas to total pressure is $\beta\ll1$ and the first adiabatic exponent can be evaluated as
\begin{equation}
\Gamma_1={4\over3}+{\beta\over6}={4\over3}+{\beta_c\over6}{\beta\over\beta_c}
\label{eq-gamma1}\end{equation}
We take the central value $\beta_c\ll1$ as the parameter for the departures from the Eddington limit,
like $\sigma\ll1$ represents the departures from the Newtonian limit.
For post-Newtonian stars near the Eddington limit, integrals (\ref{eq-I1adim}-\ref{eq-I2adim}-\ref{eq-I3adim}-\ref{eq-I4adim})
can be developped in these two parameters, neglecting second-order terms, $\mathcal{O}(\sigma^2)$ or $\mathcal{O}(\beta_c\sigma)$.
Notice that, once used equation (\ref{eq-gamma1}), each integral has a global $\sigma$ or $\beta_c$ factor
except $I_2$, where the $\sigma$ factor has to be extracted from the derivatives of the metric.
These derivatives are taken from equations (\ref{eq-a}-\ref{eq-b}), and we need only $\mathcal{O}(\sigma)$ terms:
\begin{eqnarray}
{\dif a\over\dxi}&=&4\sigma{\varphi\over\xi^2}					\label{eq-da}\\
{\dif b\over\dxi}&=&-4\sigma{\varphi\over\xi^2}+4\sigma\xi\theta^3	\label{eq-db}
\end{eqnarray}
(using equation \ref{eq-continuadim}).
In this way, each integral has either a global $\sigma$ or $\beta_c$ factor,
which makes all the other relativistic corrections useless.
In particular, we can ignore all other departures from Minkowskian metric and take $e^{3a+b}=e^{3(a+b)}=1$.
Finally, the pressure gradient in $I_4$ can be eliminated with classical hydrostatic equilibrium (\ref{eq-hydroadim}).
Once removed all the second-order terms, the integrals read
\begin{eqnarray}
I_1&=&{3\over2}\beta_c{P_c\over\alpha^3}\int{\beta\over\beta_c}\psi\xi^2\dxi				\label{eq-I1adim1}\\
I_2&=&-16\sigma{P_c\over\alpha^3}\int\left(\xi\theta^3+{2\varphi\over\xi^2}\right)\psi\xi^3\dxi	\label{eq-I2adim1}\\
I_3&=&8\sigma{P_c\over\alpha^3}\int\psi\theta^3\xi^4\dxi								\label{eq-I3adim1}\\
I_4&=&-16\sigma{P_c\over\alpha^3}\int\varphi^2\theta^3\dxi							\label{eq-I4adim1}
\end{eqnarray}
and their sum is
\begin{eqnarray}
&&{2\alpha^3\over3P_c}\sum_{i=1}^{4}I_i
=\beta_c\int{\beta\over\beta_c}\psi\xi^2\dxi										\\&&\qquad\qquad\qquad
-{16\over3}\sigma\left(\ \int\xi^4\psi\theta^3\dxi+2\int\varphi^2\theta^3\dxi					\right.\\
&&\left.\qquad\qquad\qquad\qquad\qquad\qquad\qquad\quad+4\int\xi\psi\varphi\dxi\ \right)	\label{eq-somme}
\end{eqnarray}
The first integral in the $\sigma$ term can be expressed as a function of the other two, using
(i) the continuity equation (\ref{eq-continuadim}),
(ii) an integration by parts (boundary terms vanish),
(iii) hydrostatic equilibrium (\ref{eq-hydroadim}):
\begin{eqnarray}
\int\xi^4\psi\theta^3\dxi	\stackrel{\rm (i)}{=}\int\xi^2\psi{\dphi\over\dxi}\dxi
					\stackrel{\rm (ii)}{=}-\int\varphi{\dif\over\dxi}\left(\xi^2\psi\right)\dxi	\\
					\stackrel{\rm (iii)}{=}-2\int\xi\psi\varphi\dxi+4\int\varphi^2\theta^3\dxi
\end{eqnarray}
As a consequence, the GR instability occurs when
\begin{eqnarray}
{\beta_c\over\sigma}<{32\over3}{\int\xi\psi\varphi\dxi+3\int\varphi^2\theta^3\dxi\over\int{\beta\over\beta_c}\psi\xi^2\dxi}=:\crit
\label{eq-instadim}
\end{eqnarray}
Since we extracted already the post-Newtonian correction through $\sigma$,
the integrals in equation (\ref{eq-instadim}) can be evaluated numerically from Newtonian structures ($\sigma\to0$),
either numerical or analytical.
Departures from the Eddington limit are already included in the factor $\beta_c$ and the function $\beta/\beta_c$, i.e. by the effect of gas pressure,
and one does not need to account for corrections due to the entropy of gas, which justifies a choice of the index $n=3$ in the core (section \ref{sec-sms}).
Thus, it allows to capture consistantly the GR instability on the classical hylotropic models of section \ref{sec-hylo}.
Distinctions between relativistic- and rest-mass in the hydrostatic structures are meaningless.

For given dimensionless structures, determined by the functions $\theta(\xi)$, $\varphi(\xi)$, $\psi(\xi)$ and $\beta(\xi)/\beta_c$,
equation (\ref{eq-instadim}) gives a critical ratio of $\beta_c$ and $\sigma$.
On the other hand, by definition of $\beta_c$ (\ref{eq-eos}-\ref{eq-beta}) and $\sigma$ (\ref{eq-sigma}),
the actual product of these two quantities is directly given by the central temperature $T_c$ and the chemical composition:
\begin{equation}
\beta_c\sigma={\kb T_c\over\mu\mh c^2}
\label{eq-betasigma}\end{equation}
Thus, for given dimensionless structures, the properties of the star at the limit of stability are fully determined by central temperature and chemical composition,
since $\beta_c$ and $\sigma$ are both defined.

\section{Results}
\label{sec-res}

\begin{figure}\begin{center}
\includegraphics[width=.47\textwidth]{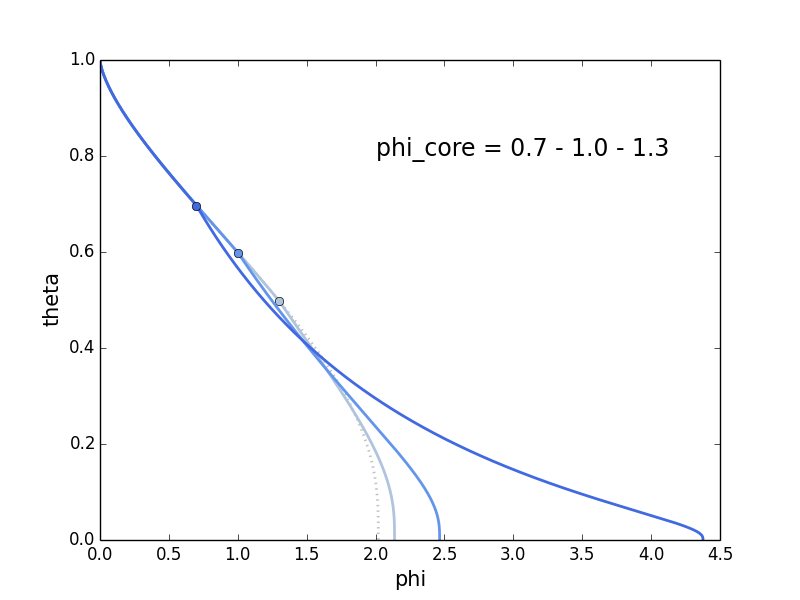}
\caption{Hylotropic structures built according to section \ref{sec-hylo},
for the indicated values of \phicore, the dimensionless mass of the core.
The dotted line is the polytrope used for the core, and a circle indicates \phicore\ for the various models.}
\label{fig-hylo}\end{center}\end{figure}

\begin{figure}\begin{center}
\includegraphics[width=.47\textwidth]{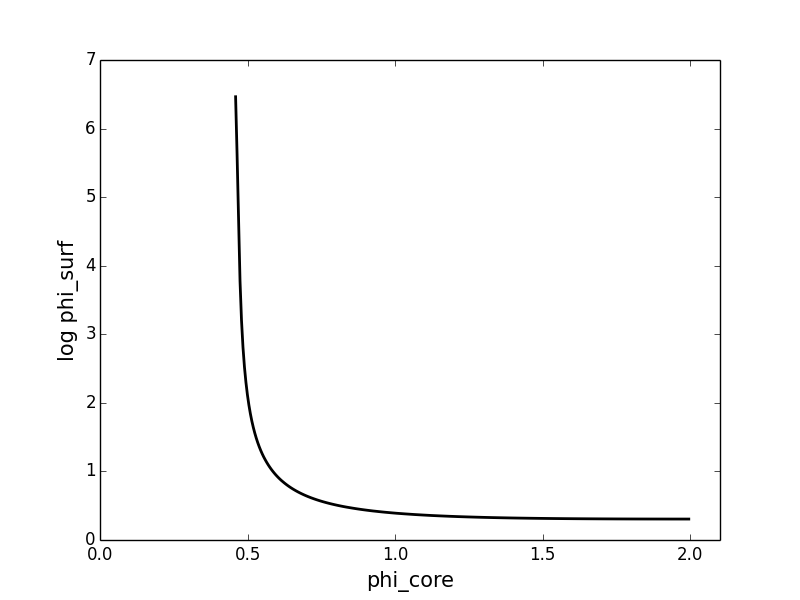}
\caption{Total dimensionless mass as a function of the dimensionless mass of the core,
for the series of hylotropic structures built according to section \ref{sec-hylo}.}
\label{fig-phi}\end{center}\end{figure}

From the full series of hylotropic structures built according to section (\ref{sec-hylo}),
three are shown in figure \ref{fig-hylo}, together with the polytrope.
As noticed by \cite{begelman2010}, a {\it de}crease in \phicore\ translates into an {\it in}crease in \phisurf, the total dimensionless mass.
This is because the pressure gradient is flatter with the hylotropic law than with the polytropic one, due to the dependence on the mass-coordinate.
Thus, larger masses are reached before pressure and density vanish.
To show this fact more clearly, we plot \phisurf\ as a function of \phicore\ for the full series of structures in figure \ref{fig-phi}
(notice the logarithmic scale in the $y$-axis).
The increase remains moderate down to $\phicore\gtrsim1$,
but then \phisurf\ gains orders of magnitude to reach millions when $\phicore\to\phimin$.
The mass fraction of the core as a function of \xicore\ in \cite{begelman2010}, is well reproduced by our series of structures.

\begin{figure}\begin{center}
\includegraphics[width=.47\textwidth]{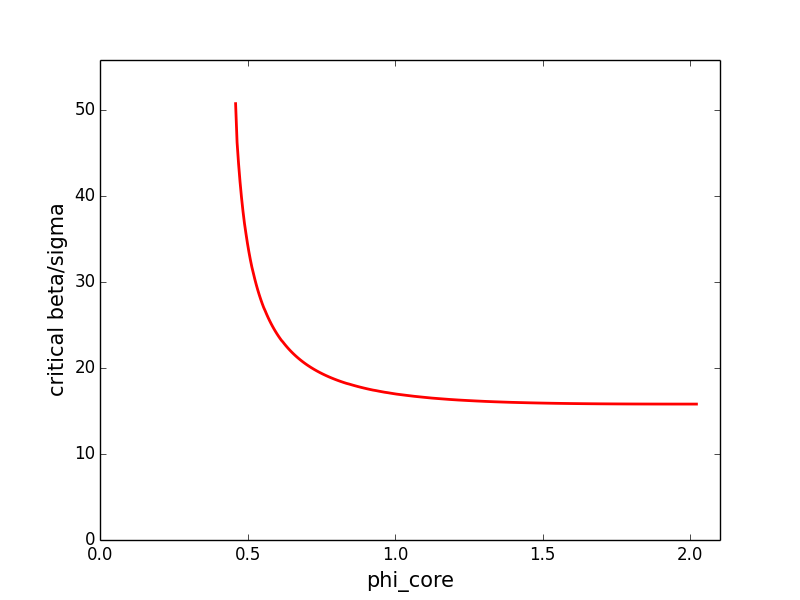}
\caption{Critical ratio \crit\ of equation (\ref{eq-instadim})
for the series of hylotropic structures built according to section \ref{sec-hylo}.}
\label{fig-beta}\end{center}\end{figure}

For this series of structures, we inject the functions $\theta(\xi)$, $\varphi(\xi)$, $\psi(\xi)$ and $\beta(\xi)/\beta_c$
(equations \ref{eq-hyloadim}-\ref{eq-betaadim}) into the stability condition (\ref{eq-instadim}).
It gives the critical ratio \crit\ as a function of \phicore, which is shown in figure \ref{fig-beta}.
In the polytropic limit $\phicore\to\phimax$,
we find a value 15.8 that reproduces well the numerical tables of \cite{chandrasekhar1964} and \cite{tooper1964b}.
When $\phicore<\phisurf$, the critical ratio departs towards larger values.
It could be interpreted naively as a destabilising effect, but it reflects only the increase in \phisurf\ that results from the decrease in \phicore.
In other words, we compare objects of different masses.
We can see already that, when the total dimensionless mass gains orders of magnitudes (figure \ref{fig-phi}),
the critical ratio \crit\ grows only by a factor of a few, which suggests that hylotropes with low \phicore\ remain stable up to larger masses than polytropes.

In order to quantify this effect, we fix the product $\beta_c\sigma$ according to equation (\ref{eq-betasigma}),
and impose critical conditions $\beta_c/\sigma=\crit$.
The central temperature of Pop III SMSs is always comprised in the interval $1.5-2\times10^8$ K
and the chemical composition does not depart significantly from $\mu=0.6$ \citep{hosokawa2013,haemmerle2018a}.
With these values in equation (\ref{eq-betasigma}) and the critical condition, $\beta_c$ is determined,
which gives the mass-scale through equation (\ref{eq-mscale}).
It allows to express the mass of the star and that of the core in solar units.

\begin{figure}\begin{center}
\includegraphics[width=.49\textwidth]{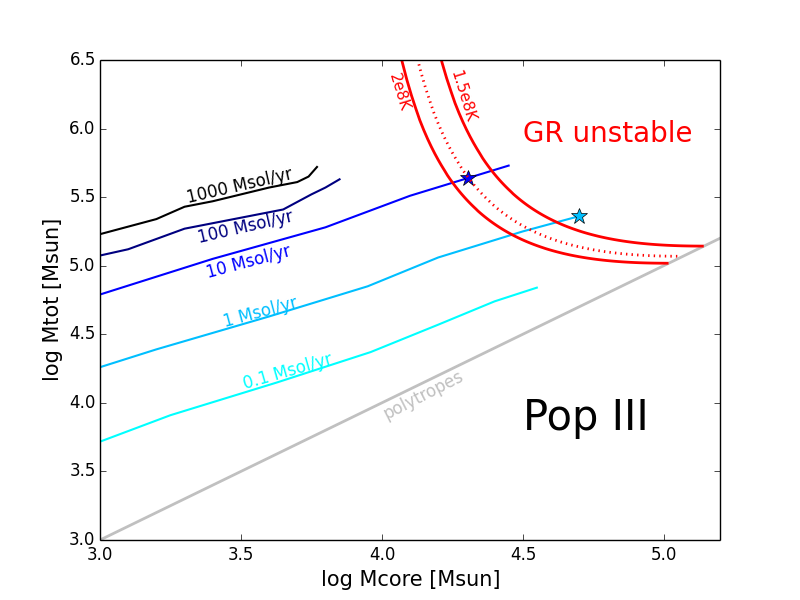}
\caption{Total mass of the star vs. mass of the convective core in the Pop III case.
The solid red lines indicate the limit of stability for hylotropic models with $\mu=0.6$ and $T_c=1.5-2\times10^8$ K.
The \gva\ models of \cite{haemmerle2018a,haemmerle2019c} at zero metallicity are plotted with the indicated rates.
The identity corresponds to polytropes of different masses, and is shown as a grey line.
The final masses obtained in \cite{haemmerle2020b} are marked by star-like symbols.
The dotted red line is the limit of stability for hylotropes with $\mu=0.6$
and the exact central temperature of the \gva\ model at 10 \Mpy\ at the onset of instability ($T_c=1.78\times10^8$ K).}
\label{fig-mmcore}\end{center}\end{figure}

The result is shown in figure \ref{fig-mmcore}.
We see that the changes in central temperature over the narrow range of relevant values does not change significantly the stability limit.
For polytropic structures, the maximum mass consistant with stability is $1-1.5\times10^5$~\Ms, in agreement with \cite{woods2020}.
But when the mass fraction of the convective core decreases, we see that the limit of stability moves towards larger masses.
In other words, the hylotropic law in the envelope has a stabilising effect against adiabatic pulsations,
as it decreases the compactness.
The effect becomes extremely strong when the mass of the core stays as low as $\sim10^4$~\Ms.
In this case, masses in excess of $10^6$~\Ms\ could be reached.
It results from the sensitive dependence of the total mass on the core mass when $\phicore<1$
while the critical ratio \crit\ does not change much, as noticed above.
Thus, the mass of the convective core appears as a key quantity,
and a star hardly reaches instability if its convective core does not exceed $\sim10^4$ \Ms.
It is only for total masses $\gtrsim10^7$ \Ms\ (not shown on the figure) that the stability limit moves to $\Mcore<10^4$~\Ms.
At the limit of gravitational binding ($\phicore\to\phimin$, equation \ref{eq-phiminmax}),
we obtain the largest mass consistant with stable equilibrium, which is $\sim5\times10^{10}$ \Ms, with a core of $\sim5000$ \Ms\
(i.e. a core mass fraction of $10^{-7}$).

\begin{figure}\begin{center}
\includegraphics[width=.49\textwidth]{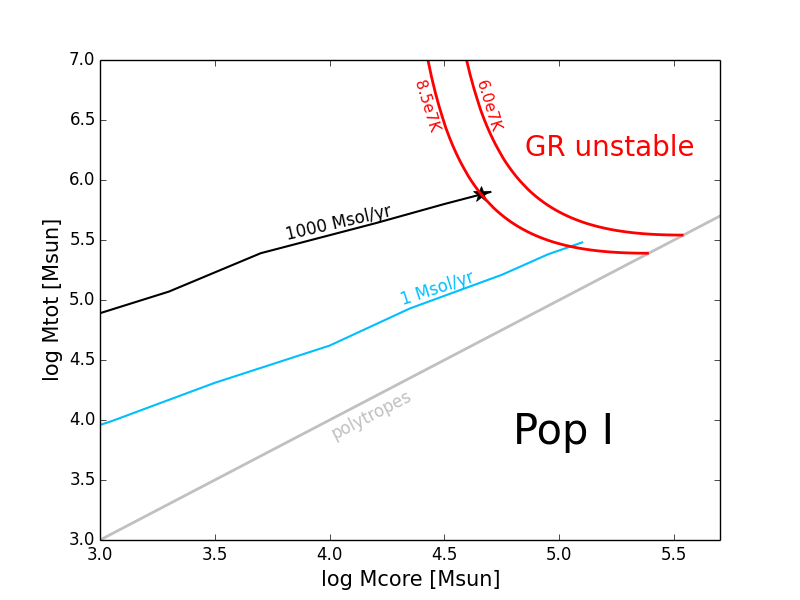}
\caption{Same as figure \ref{fig-mmcore} in the Pop I case.}
\label{fig-mmsol}\end{center}\end{figure}

In the Pop I case, shown in figure \ref{fig-mmsol}, the lower central temperatures shifts the stability limit towards higher masses.
It allows to reach $\sim3-5\times10^5$ \Ms\ in the polytropic limit,
and the typical core's mass required for GR instability moves up to $\sim3-5\times10^4$~\Ms.

\section{Discussion}
\label{sec-dis}

The reliability of the hylotropic models can be tested by a comparison with the results of the \gva\ models
\citep{haemmerle2018a,haemmerle2019c,haemmerle2020b}.
The entropy profiles of these models suggest that hylotropes are relevant for rates $\gtrsim100$~\Mpy\ (section \ref{sec-sms}).
The \gva\ tracks are displayed in figure \ref{fig-mmcore} for zero metallicity,
with their collapse points indicated as star-like symbols.
The comparison can be made only for 1 and 10 \Mpy, since the other models are still far from instability.
We see that the last stable model at 10 \Mpy\ lies in the critical band of hylotropic models with $T_c=1.5-2\times10^8$ K,
while that at 1 \Mpy\ overgoes slightly the limit.
To refine the comparison, we plot as a red dotted line the hylotropic limit
with the exact central temperature of the \gva\ model at 10 \Mpy\ at the onset point of instability ($T_c=1.78\times10^8$ K).
We see that the matching is perfect.
Departures from hylotropy in the entropy profile of this model (figure \ref{fig-genec10}) occur only near the surface,
where the density is low, which implies negligible changes in the GR corrections.
It is only for rates $\leq1$ \Mpy\ that the entropy profiles depart from hylotropy in the inner envelope (figure \ref{fig-genec1}).
Notice that the central temperature of the last stable model at 1~\Mpy\ is very close to $1.5\times10^8$ K,
so that the upper red curve gives the hylotropic limit relevant for this model.
Thus, even at this rate, the hylotropic limit is only exceeded by $\sim0.1$ dex in the $(M,\Mcore)$ diagram.
An extrapolation of the curve at 0.1 \Mpy\ suggests that this model, if it ever reaches the GR instability during H-burning,
does it as a fully relaxed, polytropic star.
Overall, it seems hard to find a rate for which the hylotropic limit would be exceeded significantly.
For rates $>1000$ \Mpy, an extrapolation of the tracks suggests final masses in excess of $10^6$ \Ms.

The \gva\ tracks at solar metallicity are plotted in figure \ref{fig-mmsol}, for the two available rates 1 and 1000 \Mpy.
The central temperatures that have been used for the hylotropic models
correspond to that of the last stable structure of the 1000 \Mpy\ run ($8.5\times10^7$ K)
and of the last model of the 1 \Mpy\ run ($6\times10^7$ K), still far from instability.
We see that the larger convective core of Pop I models cancels the effect of the lower central temperature,
as the tracks are shifted with the stability limit.
As expected, the stability limit given by the hylotropic models with the relevant central temperature
crosses the 1000 \Mpy\ track at the exact point of instability obtained in \cite{haemmerle2020b},
which corresponds to a mass of 759 000 \Ms.
The comparison cannot be done for lower rates due to the lack of models.
As in the Pop III case, masses $>10^6$~\Ms\ require rates $>1000$~\Mpy.

These results confirm that spherical SMSs forming in atomically cooled haloes cannot reach masses in excess of 500~000~\Ms,
while masses in excess of $10^6$ \Ms\ could be reached in the galaxy merger scenario (section \ref{sec-in}),
provided accretion rates $\gtrsim1000$ \Mpy.
Actually, such rates appear as a necessity for massive black hole formation through Pop I galaxy mergers,
since at solar metallicity masses $\gtrsim10^6$ \Ms\ are required to avoid thermonuclear explosion during the collapse \citep{montero2012}.
In this respect, the final mass of the black hole progenitor ranges in distinct intervals in the various versions of direct collapse:
$\lesssim500\ 000$ \Ms\ for atomically cooled haloes; $\gtrsim10^6$~\Ms\ for Pop I galaxy mergers.

\section{Summary and conclusions}
\label{sec-out}

We have estimated the final masses of the progenitors of massive black hole seeds in the various versions of direct collapse,
on the basis of hylotropic models built according to \cite{begelman2010}.
By a comparison with the \gva\ models accounting for full stellar evolution \citep{haemmerle2018a,haemmerle2019c,haemmerle2020b},
we have shown that hylotropic models predict the final masses with high accuracy for accretion rates $\geq10$ \Mpy,
and remains a very good approximation even for lower rates.

The hylotropic law in the envelope has a stabilising effect against adiabatic perturbations to equilibrium,
as it decreases the compactness and thus reduces the destabilising GR effects.
The mass of the convective core is decisive for the stability of the star.
The lower is the core's mass, the larger is the total mass consistent with stable equilibrium (figures \ref{fig-mmcore}-\ref{fig-mmsol}).
Typical conditions for GR instability are a total mass $\gtrsim10^5$~\Ms\ with a core mass $\gtrsim10^4$~\Ms.
For a core mass as low as $\lesssim10^4$~\Ms, total masses in excess of $10^6-10^7$~\Ms\ remain consistent with stability.

These results confirm that spherical SMSs forming in atomically cooled haloes (Pop III, $\dm\lesssim10$ \Mpy) have masses $\lesssim500\ 000$ \Ms.
SMSs formed by accretion at rates $\gtrsim1000$~\Mpy\ could reach masses $\gtrsim10^6$ \Ms,
which is a required condition for massive black hole formation in Pop I galaxy mergers.
It implies that the final masses of the progenitors of massive black hole seeds range in distinct intervals in the various versions of direct collapse.

\begin{acknowledgements}
LH has received funding from the European Research Council (ERC) under the European Union's Horizon 2020 research and innovation programme
(grant agreement No 833925, project STAREX).
\end{acknowledgements}

\bibliographystyle{aa}
\bibliography{bib}

\end{document}